# On the giant supercluster binary-like system formed by the Corona Borealis and Abell 2142

**Giovanni C. Baiesi Pillastrini**[1*]

**ABSTRACT**

The recent hypothesis of a giant supercluster binary-like structure formed by the Corona Borealis and its close companion Abell 2142 (supercluster) belongs to a little known area of investigation as the dynamics of gravitationally interacting galaxy supercluster pairs. From the observational point of view this structure approximates the configuration of a binary-like system in linear orbit interconnected by a huge filamentary structure which, if confirmed, it would be the first case to date observed at $z \geq 0.07$. Given the importance to disentangle this issue, a follow-up analysis has been performed on the region constrained by the common envelop of the two superclusters in order to search for new hints to confirm their mutual gravitational interaction. Observational signatures of that interaction have been found mapping the inner peculiar motions where the observed negative peculiar velocities measured within the A2142 (supercluster) region suggest a general matter flow toward the Corona Borealis supercluster. Besides, analyzing the effects on both superclusters due to the mutual impact of the external tidal forces, turns out that their inner dynamics remain unperturbed up to the turnaround radii. Outside, where the binding forces are overlapped by the tidal ones, the outskirts of both superclusters should be unstable and subject to fragmentation. Such a scenario indicates that both superclusters interact with comparable and reciprocal tidal perturbations leaving the whole system in a substantial dynamical equilibrium. The origin of such a dynamical dichotomy would be explained either by a much more massive Corona Borealis supercluster than that estimated in the present work or by a selection effect biasing the small sample of peculiar velocities due to the remoteness of the system worsened by the large uncertainty on their measurements.

[1] *Independent researcher*
[*] *permanent address: via Pizzardi, 13 - 40138 Bologna - Italy - email: gcbp@micso.net*

# 1. INTRODUCTION

Galaxy superclusters are the largest associations of galaxy groups and clusters in the Universe. Unlike galaxy clusters, superclusters are not virialised and rarely reach an equilibrium configuration. Therefore, superclusters may be defined as condensations of galaxy groups and clusters observed at different dynamical phases of evolution characterized by a virialized core and an external shell constrained by the turnaround radius which, eventually, will collapse in the future in a more homogeneous structure if the binding force will prevail on the Hubble expansion. Outside this "main body", the supercluster outskirt is limited by the so-called zero-gravity surface characterized by lower density of sparse objects which may or may not collapse in the future towards the main body. From the observations, we have learned that superclusters are isolated systems where their formation happened in a single event and rarely via merging of two or more superclusters. On the other hand, they typically reside only a few supercluster radii from one another whereby likely suffer from mutual tidal disturbances. The first catalog which identify these massive objects was compiled by Abell (1961) using the density enhancement in space from the Catalog of Galaxy Clusters (Abell 1958). With a similar methodology, were built the catalogs of Bahcall and Soneira (1984), the all-sky supercluster catalog of Abell-ACO clusters by Zucca et al. (1993) and the New Catalog of Kalinkov and Kuneva 1995. Afterwards, a new generation of supercluster catalogs were constructed with more accurate and complete data sets combined with new methodologies of clustering analysis providing more insight on the nature, extension, membership and distribution ( Einasto et al. 1997, Einasto et al. 2006; Luparello et al. 2011, Liivamagi et al. 2012, Nadathur and Hotchkiss 2013, Chow-Martinez et al. 2014). When Bahcall and Soneira (1984) advanced the hypothesis that the Corona Borealis and A2142 superclusters are likely evolving to a future collapse to form a singular, extended structure, they probably predicted the first event to date of gravitationally interacting superclusters at intermediate redshift. Also Luparello et al. (2011), making their own supercluster catalog achieved the same conclusion using a smoothed luminosity density map derived from the SDSS-DR7 galaxy survey. However, since then, no follow-up study has been performed to investigate the issue from a dynamical point of view. Stimulated by these previous observations, Baiesi Pillastrini (2016, BP16 hereafter), using a complete sample of galaxy groups and clusters lying in the dense region of sky around the Corona Borealis supercluster, searched for new hints, if any, outlined by gravitational features unrevealed by previous studies. Applying a new clustering algorithm based on the identification of the deepest potential wells in the potential distribution provided by the underlying mass distribution of the sample, three massive superclusters were identified as the well-known Corona Borealis, A2142 and Virgo-Serpent, all encircled in a region of ~ 150 $h^{-1}Mpc$ diameter. In particular, the system formed by the Corona Borealis and A2142 (hereafter CBSCL and A2142SCL, respectively) was outlined by the common isodensity contours of the deepest potential wells confirming the old hypothesis of a single, interacting structure. However, the most interesting discovery was the huge filamentary structure of galaxy groups joining both superclusters configuring a typical configuration of a gravitationally interacting *binary-like system*. Furthermore, a recent study of Kopilova and Kopilov (2017, KK17 hereafter), analyzing the peculiar motions of galaxy groups and clusters in the CBSCL region, found that the massive core of the A2142SCL *i.e.* the cluster A2142 (Abell 1958), holds a peculiar motion towards the CBSCL, which is a direct confirmation of a strong mutual gravitational interaction.

Given the importance of the issue in the context of the formation of large scale structures, a follow-up analysis is performed with the aim of finding more observational features to better understand the dynamics and the degree of interaction of the whole system. Our strategy invokes the dynamics of interacting objects as a result of cumulative gravitational forces generated by the mass distribution and, where the observed peculiar motions should closely reflect the dynamical evolution of the systems. Since our system should be still in the quasi-linear regime of structure formation we expect a close correspondence between the observed mass distribution and peculiar motions or, in other words, a direct causal connection between gravitational forces and the corresponding peculiar velocities where the source of peculiar motions can be identified (Tully et al. 2014; Pomarède et al. 2015). The first step of the present study is devoted to the revision of the BP16's results following a different approach to re-define the basic parameters as mass and extension of the two superclusters associated to each identified characteristic evolutionary state (virialized, turnaround, future collapse and zero-gravity). In the second step, the inner dynamics is traced by analyzing the peculiar motions inside the common envelop. Finally, features of disruptive effects, if any, due to the impact of external tidal fields on each supercluster structure are searched for assessing the degree of gravitational influence between the two superclusters and if one of them may or may not dominate gravitationally the other.

The paper is organized as follows: in Sect. 2, basic parameters for both superclusters are revisited as a function of each evolutionary state; in Sect.3, peculiar motions inside the common envelop are analyzed; in Sect. 4, the dynamics of the whole system is studied looking for tidal effects due to the mutual gravitational interaction. The methodological approach is briefly described and applied to a complete volume-limited sample of galaxy groups and clusters filling the studied region. Then, in Sect.5, concluding remarks are drawn.

# 2. REVISITING THE BP16' RESULTS

## 2.1. Importing dataset from BP16

The BP16's dataset consists of a complete volume-limited sample of galaxy groups and clusters extracted from the Catalog compiled by Tempel et al. (2014, T14 hereafter). The sample covers an extension of about 80 $h^{-1}Mpc$ radius around each supercluster reporting the following parameters of the T14 catalog: J2000 equatorial coordinates of the center as the origin, comoving distance in $h^{-1}Mpc$ (CMB-corrected), estimated dynamical mass (assuming NFW profile) in solar mass unit. To each object, the magnitude of the local gravitational potential in $10^6 \, h(km/s)^2$ unit determined by BP16 (see BP16 for detailed description) has been added. In the present study we assume: $H_0 = 100 \, h \, km \, s^{-1} \, Mpc^{-1}$, $\Omega_m = .27$ and $\Omega_\Lambda = .73$ according to the cosmological parameters used for the BP16's dataset. For the re-analysis restricted to the common envelop of the two superclusters, a subsample of 170 groups and clusters has been selected within the region delimited by: $223° < ra < 245°$, $18° < dec < 40°$, $200 \leq$ comoving distance $\leq 280 \, h^{-1}Mpc$. An additional cut-off has been applied to select only objects with a measured local potential well deeper than $-0.4 \times 10^6 \, h(km/s)^2$. This further selection enables to outline the "skeleton" of the system formed by objects where their spatial locations trace the deepest potential wells highlighting the densest main body with respect to peripheral meaningless details. In Fig. 1, a 3D visualization in Cartesian coordinates of the whole system shows its geometrical structure shaped by isosurface density contours of the volume data. The coordinate conversion from Equatorial (ra, dec) to Cartesian (x, y, z) has been obtained using: $x = d \, cos(ra) \, cos(dec)$, $y = d \, sin(ra) \, cos(dec)$ and $z = d \, sin(dec)$. The density isosurfaces are drawn by a 3D kernel density function (kde3D) as part of a R-code which returns a three-dimensional array of estimated density values using 100 grid points with a bandwidth of 3.8 $h^{-1}Mpc$ and displays isosurface contours at a certain level (Feng and Tierney 2008). The main components of the binary-like system appear well defined.

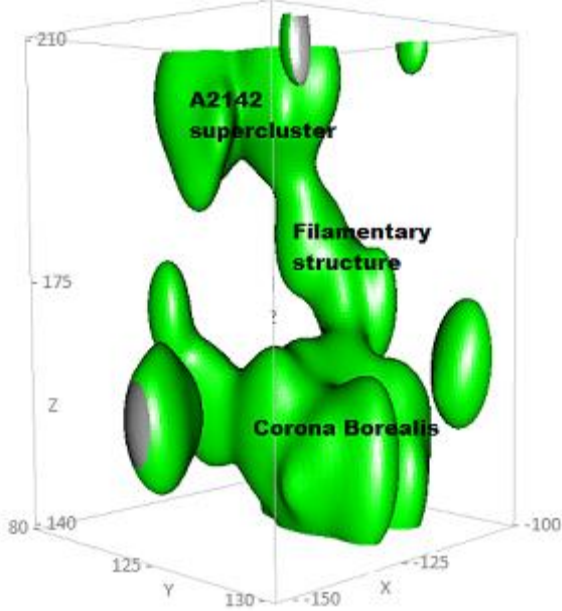

**Fig. 1** shows the 3D density isosurface contours in Cartesian coordinates of the Corona Borealis-A2142 supercluster system as a geometrical structure obtained by volume data selected as a function of the local gravitational potentials deeper than $-0.4 \times 10^6 \, h(km/s)^2$ thus forming the "skeleton" of the clustered structure. Both superclusters and the joining filamentary structure are apparent. Distances are in $h^{-1}Mpc$ unit.

**2.2. Supercluster basic parameters associated to each characteristic evolutionary state**

In BP16, the estimated masses and radii for both superclusters were approximated by applying the maximum turnaround radius-mass relation predicted by the spherical collapse model in the framework of the ΛCDM cosmology (Dunner et al. 2006, Chernin et al. 2009, Merafina et al. 2014, Pavlidou and Tomaras 2014). However this relation provides only the *maximum mass* at the *maximum turnaround* constrained by the model. It cannot provide any other information about mass and radius as a function of the evolutionary states *i.e.* which part of the supercluster is virialized, collapsing (at turnaround) or may collapse in the future or at the zero-gravity surface. To obtain a rigorous definition of each supercluster mass and radius associated to the corresponding evolutionary state, the *characteristic* density contrast criteria derived by Chon et al. (2015) and Gramann et al.(2015) are assumed. These criteria were determined via numerical simulations in the context of the spherical collapse model applied to different parametrizations of the ΛCDM cosmological model and evolutionary states. Values defining the characteristic density contrast associated to each evolutionary state and consistent with the parametrization of our fiducial cosmological model ($H_0 = 100\ h\ km\ s^{-1}\ Mpc^{-1}$, $\Omega_m = .27$) are shown in Table 1 of Gramann et al. (2015). They are obtained by the density ratio $\Delta_c = \rho/\rho_m$ of the overdensity $\rho$ to the background density $\rho_m$ that is, $\Delta_c$ = 360 (virial), 13.1 (turnaround), 8.73 (future collapse) and 5.41 (zero gravity). To pursue this goal, the revision of the BP16's results are performed with the following procedure:

1) the estimation of the supercluster mass $M(R)$ as a function of each associated evolutionary state is provided by the *dynamical mass summation* of individual galaxy groups and clusters inside a certain radius $R$ corrected by a bias factor of 1.83. Such a correction was derived by Chon et al. (2014) adopting a scaling relation obtained from cosmological N-body simulations in order to define the bias when the dynamical mass summation for group/cluster belonging to the supercluster is converted in the total supercluster mass. The accuracy of this bias factor was confirmed by Einasto et al. (2015) analyzing the mass distribution within the A2142SCL;

2) the method to estimate the mass and radius for each characteristic evolutionary state is quite simple since superclusters are expected to fit positions of maximum density contrast inside the distribution of the sampled galaxy groups and clusters. Starting from the center of each supercluster initially assumed at the position of its most massive cluster *i.e.* A2065 for the CBSCL and A2142 (cluster) for A2142SCL, the equality $\Delta(R) = \Delta_c$ must be satisfied knowing that the density contrast inside the volume $V(R)$ defined by the test-radius $R$ is given by $\Delta(R) = \rho(R)/\rho_m$ where $\rho(R) = M(R)/V(R)$; $\rho_m = 3\ \Omega_m\ H_0^2\ /8\pi G$ and $M(R)$ is the mass inside $V(R)$. The calculation is repeated iteratively for $n$ concentric spheres with increasing test-radius $R$ until $\Delta(R) \rightarrow \Delta_c$ for the corresponding evolutionary state. After each step, the new center of mass position is recalculated within $V(R)$. Then, the process is repeated checking if $\Delta(R)$ exceeds $\Delta_c$. At the equality $\Delta(R) \approx \Delta_c$ we obtain the final position of the center of mass as well as the final $\Delta(R)$ and $M(R)$ as a function of $R$ (the final $R$ and $M$ are labeled as $R_\Delta$ and $M_\Delta$). The numerical process is performed for both superclusters until the basic parameters associated to each characteristic evolutionary state are defined.

The results for the CBSCL and A2142SCL are listed in Table 1 where in Col.(1) the considered characteristic evolutionary state is labeled as: VIR→virial, TA→turnaround, FC→future collapse, ZG→zero-gravity; Col.(2) the equatorial coord. (J2000) in degree of the mass center; Col.(3) comoving distance of the mass center; Col.(4) the estimated mass $M_\Delta$; Col.(5) the density contrast $\Delta$; Col.(6) the characteristic density contrast $\Delta_c$ of Gramann et al. 2015; Col.(7) radius $R_\Delta$ when $\Delta(R) \approx \Delta_c$; Col.(8) the tidal radius as defined in Sect. 4.1 and Col.(9) the differentials $R_\Delta - R_{Tidal}$ as defined in Sect. 4.4.

| TABLE 1 - Supercluster physical parameters as a function of the characteristic evolutionary states | | | | | | | | |
|---|---|---|---|---|---|---|---|---|
| Evolutionary state | Mass center position RA/Dec ° | Mass center distance $h^{-1}Mpc$ | $M_\Delta$ $10^{15}h^{-1}M_\odot$ | $\Delta$ | $\Delta_c$ | $R_\Delta$ $h^{-1}\ Mpc$ | $R_{Tidal}$ $h^{-1}\ Mpc$ | $R_\Delta - R_{tidal}$ $h^{-1}\ Mpc$ |
| (1) | (2) | (3) | (4) | (5) | (6) | (7) | (8) | (9) |
| **CORONA BOREALIS SUPERCLUSTER** | | | | | | | | |
| Cluster A2065 (starting center) | 230.7  27.8 | 213.3 | 1.53 | | | 3.2 | | |
| **VIR** | 230.7  27.8 | 213.3 | 2.8 | 366 | 360 | 2.9 | 7.3 | -4.4 |
| **TA** | 230.7  27.8 | 213.8 | 3.4 | 13.2 | 13.1 | 9.3 | 9.31 | -0.01 |
| **FC** | 230.9  27.8 | 214.5 | 4.6 | 8.73 | 8.73 | 11.4 | 10.2 | +1.2 |
| **ZG** | 230.9  27.8 | 214.8 | 5.3 | 5.42 | 5.41 | 14.6 | 12 | +2.6 |
| **A2142 SUPERCLUSTER** | | | | | | | | |
| Cluster A2142 (starting center) | 239.5  27.3 | 265 | 0.907 | | | 2.38 | | |
| **VIR** | 239.5  27.3 | 264.6 | 1.66 | 380 | 360 | 2.4 | 6.9 | -4.5 |
| **TA** | 239.5  27.0 | 263.9 | 3.3 | 13.3 | 13.1 | 9.2 | 9.5 | -0.3 |
| **FC** | 239.7  27.0 | 263.6 | 3.34 | 8.75 | 8.73 | 10.7 | 10.23 | +0.47 |
| **ZG** | 239.7  27.0 | 263.0 | 3.9 | 5.49 | 5.41 | 13.1 | 11 | +2.1 |

The comoving separation between the mass centers of the two superclusters is ~ 58.2 $h^{-1}Mpc$.

## 2.3. Uncertainties

The uncertainties on $\Delta(R)$ and $M(R)$ mainly depend on the error affecting the mass estimates. The volume-limited group/cluster catalog of T14 does not provide the fractional error associated to each mass estimate. However, two recent studies of Old et al. (2014, 2015) comparing different mass estimations using simulated mock galaxy catalogs found the estimated mass errors affecting the Catalog of T14 show ~ 50% scatter compared with the true values. Since galaxy pairs and triplets are the major sources of the evaluated total error, a reduction of the error to 30% has been obtained assigning to these objects a proxy mass given by their sample median (Einasto et al. 2015).

**2.4. Comparison with other studies**

A detailed description of comparisons with other studies was reported in BP16 (and references therein). In what follows only those parts which are relevant for the present analysis are taken into account.

**2.4.1. The CBSCL**

As expected, all mass estimates reported in Table 1- Col.(4) compared with those of Pearson et al.(2014) *i.e.* 0.6-12 x$10^{16}$ $h^{-1}M_\odot$ ), Small et al. (1997, 1998) *i.e.* 3.3-8 x $10^{16}$ $h^{-1}M_\odot$ ) and Postman et al. (1988) of 8 x $10^{15}$ $h^{-1}M_\odot$ , are systematically underestimated confirming that the dynamical mass summation method is very conservative in estimating supercluster masses (Chon et al. 2014; Einasto et al. 2015). Instead, a substantial agreement has been found for the size of the collapsing core (at turnaround) where the radius reported in Col.(7) is slightly lower than ~ 12.8 $h^{-1}Mpc$ estimated by Pearson et al.(2014) and ~ 13±1.8 $h^{-1}Mpc$ by Postman et al. (1988); while it is comparable with ~ 10 $h^{-1}Mpc$ estimated by Small et al. (1997, 1998). The small offset of the center of mass toward higher comoving distance is likely due to the influence of the rich cluster A2161 close to the central cluster A2065.

**2.4.2. The A2142SCL**

The supercluster is divided into a higher-density core centered on the cluster A2142 and a lower-density outskirt region from which a straight and extended filament departs causing the offset of its center of mass toward lower comoving distances. Einasto et al. (2015) and Gramann et al. (2015), on the basis of the density contrast test found that only the high density core region of 6-8 $h^{-1}Mpc$ radius has reached the turnaround and starts to collapse, a radius lower than our of 9.2 $h^{-1}Mpc$. Again, the size of the A2142 "main body" evaluated at the zero-gravity surface of ~13 $h^{-1}Mpc$ radius agrees very well with our estimate. Our estimated mass of 3.3 x $10^{15}$ $h^{-1}M_\odot$ at turnaround radius is comparable (within the error) with their estimate of 2.9 x $10^{15}$ $h^{-1}M_\odot$ . Even if some differences due to different applied methods of estimation are evident, no significant discrepancies on the results can be claimed.

**3. PECULIAR MOTIONS**

If gravity dominates the dynamics, the distribution of peculiar motions may provide information on the dynamical state and future evolution of our system. In other words, to better understand if the CBSCL and A2142SCL are in dynamical equilibrium or expanding (outgoing) or collapsing (incoming), theory requires a detailed analysis of peculiar motions induced by internal matter flows and external gravitational forces. In the present case, we assume that measured peculiar velocities within the common envelop could be analyzed as in isolation knowing that the closer supercluster is the Virgo-Serpent (BP16) separated by ~ 97 $h^{-1}Mpc$ from the CBSCL and ~ 116 $h^{-1}Mpc$ from the A2142SCL so that a negligible external gravitational influence are expected. As anticipated in the Introduction and in Sect. 2.1., KK17 studying the peculiar motions of galaxy clusters in the CBSCL region, confirmed that the CBSCL and A2142SCL are gravitational interacting systems. They reached this conclusion since the cluster A2065 (the richest and centrally located cluster within the CBSCL) shows a negligible peculiar velocity indicating that the CBSCL core (at least) is not subject to gravitational attraction from the A2142SCL, while the more remote clusters A2019 and A2061 move with positive peculiar velocities toward the A2142SCL. On the contrary, the massive core of the A2142SCL *i.e.*, the cluster A2142, moves towards the CBSCL with a negative peculiar velocity of 1,343±510 $km\ s^{-1}$ suggesting the hypothesis of a general dragging of the whole A2142SCL towards the CBSCL designed as the dominant gravitational attractor of the system. Note that the angle between the two lines of sight passing through the centers of the two superclusters diverges of only 7.8° so that, peculiar velocities do not change significantly with respect to the measured radial ones. In order to reinforce the observations of KK17, a larger sample of measured radial peculiar velocities within the common envelop is searched for. Unfortunately, our superclusters and, in particular, the A2142SCL lie at a distance where selection effects (mainly due to the Malmquist bias) restrict the measurability of peculiar velocities ($z$ ~ 0.1) and the error affecting measurements is very large (> 100%). In literature, peculiar velocity measurements for galaxy groups can be found on the Cosmicflows-2 catalog (Tully et al. 2013, see also website http://edd.ifa.hawaii.edu). It was compiled within the framework of the Cosmicflows project (*e.g.* Courtois et al. 2011a,b; Tully & Courtois 2012) with a median error on distances of ~ 15-25% (Courtois and Tully 2012a,b; Tully et al. 2013) listing 5,224 group entries, including 4,690 of these as singles. After adaptation to our assumed cosmological parametrization, only 4 groups with a single peculiar velocity along with the cluster A2148 (where its measurement was obtained averaging data from 6 galaxy members) are found within the envelop. These radial peculiar velocities are plotted in Fig. 2 as a function of comoving distances together with those reported by KK17 for the cluster A2065, A2061, A2067, A2089 belonging to the CBSCL and, A2142, for the A2142SCL. At a first glance one can note that the regions occupied from the A2142SCL and the filamentary struc-ture show only negative peculiar velocities towards the CBSCL, a scenario expected if the CBSCL were the dominant gravitational attractor of the whole system. In particular, can be noted that the large errors affecting peculiar velocity measurements of galaxy groups extracted from the Cosmicflows-2 catalog (see black error bars in Fig.2) seem to challenge the meaning of the observed negative trend as a mass flow towards the CBSCL. However, the robust peculiar velocity measurement for the cluster A2142 obtained by KK17 (which is 2.6 times exciding the error and determined averaging the fundamental planes built for 67 early type galaxies) cannot be undervalued and may support by itself the idea of a general flow towards the CBSCL because of its gravitational dominance of the homonym supercluster associated with a predictable drag effect of the surrounding environment.

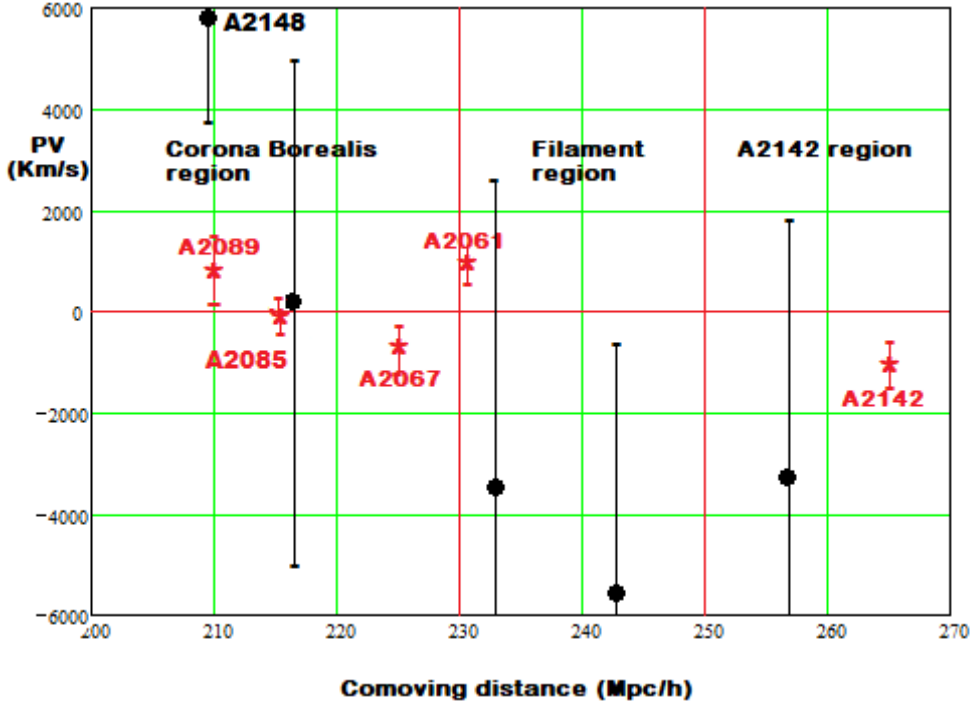

**Fig.2** Five peculiar velocity measurements of galaxy groups lying within the common envelop of the Corona Borealis-A2142 supercluster system (black asterisks/error bars) extracted from the Cosmicflows-2 catalog (see text) along with those measured by KK17 for the Abell clusters in the same region (red stars/error bars) are plotted as a function of comoving distances. The two regions covered by the filamentary structure and A2142SCL show negative peculiar motions towards the CBSCL.

A question arises at this point: can the CBSCL play the role of the attractor even if its mass and extension are comparable (within the error) with those of the A2142SCL? This balancing between supercluster masses would predict a substantial dynamical equilibrium and a random distribution in opposite directions of peculiar velocities, which is not the present case. The hypothesis of an infall of the A2142SCL towards the CBSCL stemming from our analysis is weakened by the too small sample of available peculiar velocity measurements which suffers of very large errors and likely biased by selection effects due to the remoteness of the A2142 region. Besides, given the large separation of the supercluster center of masses ( ~ 58 $h^{-1}Mpc$), the CBSCL would gravitationally dominate the whole system only if its mass should be deemed *much more* massive than that previously evaluated. In any case, we are aware of drawing conclusions from the analysis of the peculiar velocity sample within the envelop as the measurements are poorly reliable to base a claim on. In the next Section we attempt to overcome the above stalemate by performing a tidal analysis to evaluate the tidal effects due to the external mass distribution on both supercluster structures and their associated dynamics.

## 4. THE TIDAL APPROXIMATION

### 4.1. The tidal radius

What is the influence of the external gravitational forces acting on each supercluster structure? This is not an easy task since galaxy superclusters, for their complex and ill-defined dynamical state, are not astronomical bodies suitable for applications based on the conventional Newtonian mechanics *e.g.* the two-body model (which, instead, has large applications on the galaxy cluster pairs because of their approximated spherical symmetry and virial equilibrium). An alternative way of gain information about the dynamics of two objects embedded in a surrounding mass distribution can be provided by the tidal theory in the Newtonian approximation. That is, hints on the dynamics of a complex system like that being studied here, can be gained by observing the tidal effect on each body taken one at a time as a test-body. The strength of the tidal field acting on the test-body is generated by the summation of all individual strengths provided by the sampled objects assumed as point-masses in isolation within a fiducial spherical volume. This technique allows to measure the total strength of the tidal field independently from how surrounding masses are distributed including the tidal share due to objects belonging to the filamentary structure which would be neglected otherwise. From the theory, the static tidal limitation on a test-body due to the action of the external tidal field is spatially fixed by the *tidal radius* $R_{Tidal}$. It separates the inner dynamics of a test-body dominated by the binding force from the background where external tidal field dominate. Because the tidal tensor is symmetric, it could be set in orthogonal form where the eigenvalues λ measure the strength of the tide along the corresponding eigenvectors and, according to the sign, one can separate the tidal action as compressive tide if λ < 0 (favoring the infall of remote objects towards the center) or extensive if λ > 0 (favoring the stripping off ). However, we are not interested in such a detailed analysis since our focal point consists to verify if the external tidal perturbation can be disruptive causing gravitational instability on the dynamics of the test-body. This can be tested comparing $R_{Tidal}$ with the $R_Λ$ associated to each characteristic evolutionary state of both superclusters (see Table 1). It enables a stringent test to verify if the tidal perturbation is strong enough to change (or not) the inner dynamics of each supercluster.

### 4.2. Quantifying the tidal radius

It is well known that the source of the tidal force acting on a test-body is due to a time-independent gravitational potential generated by a "point mass" spherical distribution (of groups and clusters in this case) surrounding the test-body as the reference frame center and can be expressed by

$$F_{Tidal,a} \equiv -\frac{d^2\Phi_{ext}}{dR_a dR_b} R_b \equiv F_{ab} R_b \qquad (1)$$

where $R$ is the radius vector in the test-body reference frame and, in the framework of the ΛCDM cosmological model, $\Phi_{ext} = -U_g - U_Λ$ where $U_g = G \sum_{i=1}^{N} m_i / (d_i - d)$ is the *attractive component* of the potential due to gravity and $U_Λ = (2/3) Λ \sum_{i=1}^{N} (d_i - d)^2$ is the *repulsive component* of the potential due to dark energy where Λ is the cosmological constant of ~$10^{-20}$ $Mpc^{-2}$ . Since a test-body is subjects to the action of $N$ surrounding objects (assumed as point-masses) of mass $m_i (i = 1, ... N, \notin$ test body) located at position vectors $d_i$ (from the observer) within a spherical volume $V$ of fixed radius $R_V$ centered on the test-body at position vector $d$, the Newtonian tidal tensor $F_{ab}$ is a square 3×3 symmetric matrix (Hessian) given by

$$F_{ab} = \sum_{i=1}^{N} (m_i / |d_i - d|^3) \left[ (3(d_i - d)_a (d_i - d)_b / |d_i - d|^2) - \delta_{ab} \right] \qquad (2)$$

where the gravitational constant $G = 1$ and $\delta_{ab}$ is the Kronecker delta. The potential $U_Λ$ has been omitted since after the differentiation of Eq.(1) it turns out a constant $Λ/3$ term numerically negligible. It follows that the strength of the tidal force is $F_{Tidal} = |F_{aa} R_a|$ where $F_{aa}$ are the three eigenvalues corresponding to the principal axes of $F_{ab}$.
Now, using basic parameters defined for both superclusters as a function of their characteristic evolutionary states, the net tidal influence spatially defined by the tidal radius can be quantified and compared with each characteristic radius reported in Table 1. By assuming the equilibrium condition $F_{Tidal} = F_{binding} = M/R^2$ (where $M$ and $R$ are the mass and radius of the test-body reported in Table 1 as a function of each characteristic evolutionary state) the tidal radius is given by

$$R_{Tidal} = (M/F_{Tidal})^{1/2} \ . \qquad (3)$$

Obviously, the source of $F_{Tidal}$ is mainly due to the most massive perturbing counterpart of the perturbed test-body where, in this case, the CBSCL interchange the role of counterpart with the A2142SCL alternately. However, whereby hardly a supercluster can be assumed as a single point-mass as stated in the previous Section, the summation $\sum_{i=1}^{N}(m_i)$ in Eq. (2) includes all groups and clusters within $V$ and external to $R$ summated as individual masses independently from being members of the perturbing supercluster, the filamentary structure or field objects.

### 4.3. Simplifying assumptions

There is the well known problem related to the solution of the Eq.(1 ÷ 3) for infinite number of gravitating masses that is, when $(d_i - d) \to \infty$ (like the gravitational potential $\Phi_{ext}$) the tidal tensor $F_{Tidal} \to 0$ at the position of a generic test-body. To find a finite solution which overcomes this problem, an assumption is required on the form of the mass distribution in space. For our purpose, a reasonable assumption may suppose that the mass distribution inside a generic spherical volume $V$ of fixed radius $R_V$ is embedded in a *uniform background* which ensure that masses within $V$ provides a finite value of the tidal field while "external" masses with $(d_i - d) \geq R_V$ should have negligible influence on the determination of $F_{Tidal}$. In any case, $V$ should be large enough to encapsulate the major share of the tidal influence on the test body placed in its center. To establish how large $R_V$ should be assumed, we know that our superclusters approximately fill a spherical envelop with a diameter of ~ 80 $h^{-1}Mpc$ which implies that $R_V$ must be larger than that extension to prevent the so-called shot noise error as well as a variation of $F_{Tidal}$ larger than few percent on its strength. To satisfy this requirement, a fair approximation can be obtained performing a test where each supercluster is taken as a test-body so that $F_{Tidal}$ is computed as a function of increasing $R_V$ starting from its center of mass. The computations is stopped when the strength increments of $F_{Tidal}$ are lesser than 1.5% so that $R_{Tidal}$ decreases less than 1%. As expected, such a condition is satisfied when $R_V \to 80\ h^{-1}Mpc$ and $V$ includes the supercluster counterpart. Note that the data entering in Eq.(2) are extracted from the T14's Catalog.

### 4.4. Uncertainty

By knowing that errors on spectroscopic redshifts of the SDSS DR10 survey (used to construct the T14's Catalog) do not exceed a few % it is now possible a statistical evaluation of the uncertainty on the tidal radius $R_{Tidal}$. To quantify it, we apply a Monte-Carlo simulation based on the resampling technique (Andrae 2010) to a random subsample enclosed in a spherical volume of $R_V = 80\ h^{-1}Mpc$ which assumes a Gaussian error distribution of ~ 3% for comoving distances and ~ 30% for mass estimates. Then, we can now randomly sample new data points to estimate the simulated $R_{Tidal}$ at the center of the test-body. Repeating this resampling task 10,000 times, we get the distribution of the simulated data from which we can then infer the uncertainty given by the standard deviation. An estimated standard error of ~17% has been found.

### 4.5. Results

In Table 1 values of $R_{Tidal}$ computed using parameters corresponding to each supercluster evolutionary state are reported in Col.(8), while differentials given by $R_\Delta - R_{Tidal}$ are reported in Col.(9). For both the superclusters, that differentials quantify the influence of the gravitational tidal field on the stability of each characteristic dynamical state.
If our fiducial model used to quantify supercluster parameters as a function of the characteristic evolutionary state is correct, the following scenarios are expected:
A) $R_\Delta - R_{Tidal}$ is *negative*; $R_{Tidal} > R_\Delta$, in this case the two superclusters are "weakly interacting" and their internal structures remain unmodified by the external tidal perturbation.
B) $R_\Delta - R_{Tidal}$ is *positive*; $R_{Tidal} < R_\Delta$, thus the two superclusters are "strongly interacting" where the tide overlaps the binding force within each evolutionary state up to the zero-gravity surface making the whole or part of the structure dynamically unstable and where traces of tidal effects as stripped off objects may be evidenced by the distribution of peculiar velocities.
C) $R_{Tidal} \approx R_{TA}$ for both superclusters, one may suppose that they are in "interacting equilibrium". In this case, the inner dynamics constrained at the turnaround radius would not be modified by the tidal perturbation. However, objects lying in the outskirt regions constrained between $R_{TA}$ and $R_{ZG}$ may or may not collapse toward the center or even stripped off forming an unstable shell due to the overlapping of the binding and tidal forces and where inner peculiar motions would trace matter displacement.
From Table 1 one can note the similarity of the output data provided by the tidal analysis that is, both superclusters match the case C outlined above where $R_{TA} \approx R_{Tidal}$ and, $R_{FC}$ and $R_{ZG} > R_{Tidal}$. This means that the spherical outer shell between $R_{TA}$ and $R_{ZG}$ is subject to the external tidal force making the outskirt regions gravitationally unstable. Such a behavior was predicted by numerical simulations of interacting superclusters. Shaya (1984) using Montecarlo simulations, compared the distribution of external tidal forces with the binding forces in the outer regions of superclusters finding that that regions are strongly influenced by tidal fields. Exactly what is confirmed by Einasto et al. (2015) according to which the long filament extending from the main body of the A2142 supercluster would fragment into several systems in the future. Similarly, the positive peculiar velocity measured by KK17 for the cluster A2061 which moves from the border of the CBSCL towards the A2142SCL may be ascribed to the tidal interaction between them. On the other hand, the inner dynamics inside $R_{TA}$ should not undergo significant variations from tidal perturbations since the inner binding force prevails on the external tidal field which ensure that the main body of both superclusters would be gravitationally bound and likely in process of future virialization. It is interesting to note that the stringent match between the characteristic radii provided by the density contrast criteria used here in defining the supercluster evolutionary states and the tidal radii imposed by the surrounding mass distribution confirms the good agreement between theoretical predictions and observations. However, taking into account the large uncertainties affecting the parameters of Table 1, the above results and related conjectures should be taken with caution. Which could be argued with a reasonable degree of accuracy is the apparent balancing between self-gravitational forces and disruptive tidal perturbations mainly due to comparable supercluster masses (at the turnaround surface they are almost equivalent, while at the zero-gravity differ by a factor of 1.35 in favor of the CBSCL). From this scenario, one expects to observe a random distribution of negative and positive radial peculiar velocities within the common envelop and where the approaching or the moving away of the two superclusters from each other would turn out indistinguishable. On the contrary, the observed trend of negative peculiar velocity measurements in the region occupied by the A2142SCL seen in Fig. 2 seem to sug-gest a general matter flow towards the CBSCL, *a scenario incompatible with the case C*. As stated before, the dynamics of two gravitationally interacting superclusters cannot be studied applying the conventional two-body problem (see Sect. 4.1.) so that, one may use Eq. (2) and (3) to establish how much strength should have the external tidal field $F_{Tidal}$ induced by the CBSCL (and surrounding matter) to be disruptive for the whole structure of the A2142SCL. In other words, we may perform an iterative simulation based on the variation of $F_{Tidal}$ as a function of increasing proxy mass values for the CBSCL until $R_{Tidal}$ converges towards a value *lesser* than the virial radius of the A2142SCL. This occurs when the assumed proxy mass value for the CBSCL $> 10^{17}$ $h^{-1} M_\odot$ ! Note that such a value is consistent with the mass upper limit provided by Pearson et al. (2014) and discussed in Sect. 2.4.1 which may provide support to the idea of a more massive CBSCL. However, the Pearson's estimate is an extreme value of almost two orders of magnitude greater than that estimated here. We point out that on the contrary of the Pearson's estimate, our mass estimate for the CBSCL is in excellent agreement with the upper limits of 3.16 and 4.3 x $10^{15}$ $h^{-1} M_\odot$ predicted respectively by the simulations of Tinker et al.(2008) and Bolejko and Ostrowski (2018) for the most massive objects in the Universe.

## 5. CONCLUDING REMARKS

A follow-up analysis on the dynamics of the giant supercluster binary-like structure formed by the Corona Borealis supercluster coupled with the A2142 supercluster has been performed searching for new observational and gravitational hints of mutual gravitational interactions. To disentangle this issue two lines of research have been addressed focalizing the analysis on the region constrained by the common envelop of the two superclusters. Firstly, supercluster basic parameters as mass and extension have been quantified as a function of each characteristic evolutionary state using well defined characteristic density contrast criteria established in the context of the ΛCDM cosmology. The revisited basic properties and physical parameters of both superclusters have been used to analyze the dynamics of the whole system studying the effects of the external tidal impact due to the surrounding mass distribution on the inner structure of each supercluster. The result confirms that both superclusters are mutually interacting but only their outskirts are subjects to the external tide spatially fixed by tidal radii, while their main bodies remains unaffected since, at the turnaround radius, the inner binding force prevails on the external tide indicating a substantial dynamical equilibrium of the whole system. On the contrary, using peculiar motions provided by a small sample of measured radial peculiar velocities, the dynamics of the system has been reconstructed from the observational point of view. The observed negative trend emerging from the negative peculiar motions of the cluster A2142 along with few other galaxy groups seem to suggest a general infall of the A2142SCL towards the CBSCL but the large uncertainty affecting that measurements and the suspect of a strong selection bias of the peculiar velocity sample due to remoteness of the system cannot give a rigorous description of the dynamics within the common envelop. Besides, remains inexplicable how can be generated that general infall since the role of the CBSCL as the "attractor" for the A2142SCL is not supported by our tidal analysis. The conundrum of this discrepancy would be overcome when a larger and unbiased peculiar velocity sample will become available allowing a robust reconstruction of the dynamics of the whole system.